\documentclass[12pt,a4paper]{article}

\usepackage{cite}
\title{SNe Ia in Standard and Conformal Cosmological Models}

\author{A.B. Arbuzov$^{1}$, V. Shmidt$^{1,2}$, and A.F. Zakharov$^{1,3}$}

\date{}

\begin{document}

\maketitle

\begin{center} {\it
$^1$ Bogoliubov Laboratory of Theoretical Physics,
Joint Institute for Nuclear Research,
Dubna, Moscow 141980, Russia \\
$^2$ Moscow Institute of Physics and Technology,
Dolgoprudny, Moscow 141701, Russia \\
$^3$ National Research Center ``Kurchatov Institute'',
Moscow 123182, Russia
}
\end{center}

\begin{abstract} 
This article demonstrates the classical equivalence between standard and conformal cosmological models regarding the apparent luminosity of type Ia supernovae. We show that previous claims of observational discrepancies stem from differing interpretations of physical quantities when transitioning between Einstein and Jordan frames. Specifically, we establish that in the conformal picture, SNe Ia are not "standard candles" in the conventional sense because their absolute luminosity depends on redshift due to variations in local measurement units.
\end{abstract}

%\keywords{Conformal cosmology; type Ia supernovae; Einstein and Jordan frames.}

%\ccode{PACS numbers: 03.65.$-$w, 04.62.+v}

\section{Introduction}

The discovery of the accelerated expansion of the Universe based on observations of cosmological supernovae SNIa, under the assumption that they are  ''standard candles'' \cite{Perlmutter_99}, led to the conclusion about the non-zero value of the $\Lambda$-term or its possible generalization - dark energy \cite{Turner_99,Perlmutter_99b}. In recent decades,  $\Lambda$CDM has been considered the standard cosmological model because it has not been inconsistent with observational data.    
However, the analysis of modern cosmological data, based on precise measurements of the Cosmic Microwave Background (CMB) properties, Type Ia Supernova (SNe Ia) observations, and galaxy density distribution measurements using the Dark Energy Spectroscopic Instrument (DESI), poses a serious challenge to the standard $\Lambda$CDM cosmological model. For instance, the expansion rate of the Universe, estimated on the base of SNe Ia observational data~\cite{Riess:2019cxk, Riess:2021jrx}, is approximately 5$\sigma$ higher than the expansion rate estimated within the standard $\Lambda$CDM cosmology with the use of Planck collaboration CMB data (Planck+$\Lambda$CDM)~\cite{Planck:2018vyg} and Baryon Acoustic Oscillations (BAO) along with recent results from the Atacama Cosmology Telescope (ACT)~\cite{AtacamaCosmologyTelescope:2025vnj}. Furthermore, the analysis of the latest BAO results obtained by the DESI collaboration, combined with CMB and SNe Ia data, demonstrates a deviation from theoretical estimates obtained within the $\Lambda$CDM model at the 3$\sigma$ level~\cite{DESI:2024mwx}. These findings also point to a possible time dependence of dark energy~\cite{DESI:2024mwx, Cortes:2024lgw}, previously identified with the $\Lambda$-term, which leads to the conclusion that the Universe may come into a phase of decelerated expansion in the future~\cite{Son:2025rdz}. Moreover, after accounting for recently emerged indications of correlations between SNe Ia brightness and the ages of host galaxies~\cite{Son:2025rdz}, the current Universe appears to be in a state closer to decelerated expansion than to accelerated expansion. In this context, discussing the links between SNe Ia cosmological observations and alternative cosmological models is especially relevant.

The conformal cosmological model~\cite{Deser:1970hs, Dirac:1973zw, Dirac:1973gk, Behnke:2001nw, Zakharov:2010nf, Zakharov:2012, Domenech:2016yxd, Pervushin:2017zfj, Lombriser:2023afv} is one of the least radical modifications of standard cosmology. It is based on the hypothesis of the fundamentality of conformal symmetry, which is spontaneously broken to the symmetry of general covariant transformations (see, for example, the articles~\cite{Ogievetsky:1973ik,Borisov:1974bn}). With such spontaneous symmetry breaking, a Goldstone boson naturally arises, called the dilaton, which is responsible for scale (Weyl) transformations.

As is known, scalar-tensor generalizations of Einstein's General Relativity (GR), which underlies standard cosmology, can be formulated in two frames, termed the Jordan frame and the Einstein frame. In the Jordan frame, the kinetic terms of the scalar fields in the Lagrangian have a canonical form, but the Riemann curvature contains a non-constant prefactor in the form of a certain function of the scalar fields. In the Einstein frame, on the contrary, the Riemann curvature is not multiplied by scalar fields, but the kinetic terms of the latter are no longer canonical. The transition between these frames is carried out via a Weyl transformation of the metric, which will be discussed below. The standard (SC) and conformal (CC) cosmological models are linked through this transformation exactly. 

A question naturally arises regarding the physical equivalence of the Jordan and Einstein frames, which, to our knowledge, remains a subject of debate~\cite{Postma:2014vaa, Faraoni:1999hp, Barvinsky:2008ia, White:2012ya}. Most such discussions conclude that these schemes are equivalent at the classical level, but differences in predictions may appear at the quantum level (see, for example, the articles~\cite{Barvinsky:2008ia, White:2012ya}).

In a cosmological context, the question of the equivalence of Jordan and Einstein frames reduces to comparing SC and CC models~\cite{Behnke:2001nw, Zakharov:2010nf, Domenech:2016yxd, Pervushin:2017zfj, Lombriser:2023afv, Dalang:2019fma}. The physical equivalence of the latter at the classical level has been discussed and demonstrated in a number of works where specific cosmological effects are considered. These include the redshift effect, the Thomson scattering cross-section and the relic background, the Friedmann equations, and cosmological inflation~\cite{Domenech:2016yxd, Lombriser:2023afv}. All these effects, investigated within CC, received their interpretation and proved to be completely equivalent to the corresponding results in standard cosmology. Nevertheless, there are works claiming that there is a difference in observational predictions obtained in CC and SC for the apparent luminosity of type Ia supernovae. For instance, in the papers~\cite{Behnke:2001nw,Zakharov:2010nf,Dalang:2019fma} different redshift dependence factors arise for SNe Ia apparent luminosity and luminosity distances. The authors assumed that a supernova, being a standard candle in SC, would remain one within CC as well.

In this article, we show that the absolute luminosity of SNe Ia in CC depends on redshift, and, consequently, supernovae in the conformal picture are no longer standard candles, which must be taken into account when analyzing observational data. As we show below, this leads to the restoration of the equivalence of results obtained in the SC and CC models.

The article is organized as follows. In the second section, we briefly consider the conformal dilaton gravity and its connection with the General Relativity (GR) in the context of the Einstein and Jordan frames. In the third section, we discuss the conformal cosmological model and an interpretation of the redshift effect within its framework. In the fourth section, we provide a detailed analysis of the apparent luminosity of type Ia supernovae in standard and conformal cosmological models and demonstrate their physical equivalence. Finally, in the conclusion, we summarize the results of the research.

\section{Conformal Dilaton Gravity}	

The currently generally accepted standard cosmological model, $\Lambda$CDM, is formulated within the framework of GR, whose gravitational sector is given by the action\footnote{The choice of the sign in front of the Einstein-Hilbert action is dictated by useage of Lorentzian metrics of signature $(-+++)$ and matters only when gravity is coupled to matter.}
\begin{equation}\label{action_GR}
S_{\rm{GR}}[g^E_{\mu\nu}] = \frac{M_P^2}{16\pi}\int {\rm{d}}^4 x \sqrt{-g_E}\, R[g^E_{\mu\nu}],
\end{equation}
where $M_{P}$ is the Planck mass in the natural system of units $\hbar=c=1$, and $R[g_{\mu\nu}]$, as usual, denotes the scalar curvature of space-time. The added subscript ``$E$'' indicates the Einstein frame.

The shift to the Jordan frame is accomplished by means of a Weyl transformation of the metric:
\begin{equation}\label{conf_EJ}
g^E_{\mu\nu}= \Omega^2_{\psi}\, g^J_{\mu\nu}, \quad \Omega_{\psi}=\sqrt{\frac{4\pi}{3}}\frac{\psi}{M_P},
\end{equation}
which, after promoting $\psi(x)$ to a dynamical degree of freedom, leads to a scalar-tensor theory known as {\it{conformal dilaton gravity}}:~\cite{Alvarez:2014qca, Alvarez:2015mpa, Alvarez:2015ewa}
\begin{equation}\label{action_CDG}
S_{\rm{CDG}}[\psi, g^J_{\mu\nu}]= \int {\rm{d}}^4 x \sqrt{-g_J}\,\Big(\frac{1}{12} \psi^2 R[g^{J}_{\mu\nu}]+ \frac{1}{2}g^{\mu\nu}_J\nabla_{\mu}\psi\nabla_{\nu}\psi\Big).
\end{equation}
Here $\psi(x)$ is a real scalar field providing an additional invariance with respect to local Weyl scale transformations with an arbitrary positive, smooth, and dimensionless conformal factor $\Omega(x)$:
\begin{equation}\label{conf_CDG}
\psi \to \tilde{\psi}=\Omega^{-1} \psi, \quad g_{\mu\nu} \to \tilde{g}_{\mu\nu}=\Omega^2 g_{\mu\nu}.
\end{equation}
The conformal cosmological model is based on the use of action (\ref{action_CDG}).

It is useful to note that the dilaton $\psi=\psi(x)$ in the scalar-tensor model (\ref{action_CDG}) is a fictitious degree of freedom that ensures the local scale invariance of the theory, rather than an additional physical mode. This is indirectly reflected in the wrong sign in front of its kinetic term. In this sense, the wrong sign in front of $\frac{1}{2}\nabla_{\mu}\psi \nabla^{\mu}\psi$ does not mean that the theory is non-unitary. It turns out to be unitary, and all on-shell UV divergences are the same as in general relativity \cite{Alvarez:2014qca}.

Expressions (\ref{conf_EJ}) and (\ref{action_CDG}) make it possible to determine the reduced Planck mass in the Einstein and Jordan frames according to:
\begin{equation}\label{Mpl}
m_E = M_P, \quad m_J = M_P \Omega_{\psi}=\sqrt{\frac{4\pi}{3}}\psi.
\end{equation}
Scale transformations can be interpreted as local transformations of units of measurement. Indeed, if the local coordinates $x^{\mu}$ are considered dimensionless, then all the line-element dimension will be contained in the metric. In this case, the conformal transformation of the metric is essentially a local ``rescaling of rulers''. And despite the fact that in the Jordan frame the reduced Planck mass is dynamical, $m_J\propto\psi$, it is not difficult to construct a frame independent invariant:
\begin{equation}
m_{\alpha}^2\, {\rm{d}}s_{\alpha}^2={\rm{inv}}, \quad \alpha=E, J.
\end{equation}
So, the distances measured in Planck units of the corresponding frame remain invariant under a change of frame:
\begin{equation}\label{invarCC}
m_J^2 \,{\rm{d}}s^2_J=m_J^2 g^J_{\mu\nu} \,{\rm{d}} x^{\mu} {\rm{d}}x^{\nu} = m^2_E g^E_{\mu\nu} \,{\rm{d}} x^{\mu} {\rm{d}}x^{\nu} = m_E^2 \,{\rm{d}}s^2_E.
\end{equation}
This is an important remark, since now it is useful to express (measure) any observable of interest in Planck units of the corresponding scheme. The latter, in turn, will allow us to determine the conformal weight of the observable \cite{Bekenstein:1980jw, Postma:2014vaa}. For the Einstein frame, it is equal to work within the natural system of units $\hbar=c=1$. However, in the Jordan frame, the reduced Planck mass in the cosmological context depends on the scale factor (or, equivalently, on the redshift $z$), which will entail certain consequences.

\section{Cosmology}
Let us provide the discussion from the previous section with a cosmological context. The spatially flat Friedmann-Robertson-Walker metric in standard cosmology is conformally flat:
\begin{equation}
{\rm{d}}s^2_E= a(z)^2 ({\rm{d}} {\bf{x}}^2-{\rm{d}}\eta^2).
\end{equation}
Turning into the conformal picture (the Jordan frame) implies a conformal transformation to a static metric:
\begin{equation}
{\rm{d}}s^2_J= ({\rm{d}} {\bf{x}}^2-{\rm{d}}\eta^2).
\end{equation}
In the conformal dilaton gravity model (\ref{action_CDG}), the scale factor $a$ is identified with the dilaton $\psi(t)=\langle \psi(x)\rangle_{\textbf{x}}$, averaged over spatial variables by virtue of the cosmological principle (spatial homogeneity and isotropy of the Universe).

Despite the fact that the choice of the conformal cosmological model has led us to a static Universe, this does not yet mean that the redshift effect has disappeared. In standard cosmology, the redshift is due to a change in the frequency of a photon (as a consequence of the expansion of the Universe), and hence in its energy. In the conformal picture, however, the redshift is resulted from a change in the energy levels. The latter is a consequence of the dependence of the Planck mass on the scale factor. Indeed, the electron mass measured in Planck mass units of the Jordan frame is equal to:
\begin{equation}
m^J_e =m^E_e a(z) =\frac{m^E_e}{1+z}.
\end{equation}
Consequently, the Bohr radius in conformal cosmology scales as $\propto m_J^{-1}$, which means that the atomic energy levels scale inversely proportional, i.e., $\propto m_J$. We obtain time-dependent energy levels, namely:\cite{Domenech:2016yxd, Lombriser:2023afv}
\begin{equation}
E^J_n =\frac{E^E_n}{1+z}.
\end{equation}
This example alredy shows that the different frames should be called different descriptions of the same phenomenon. They do not yield new physics, but a different physical interpretation, at least at the classical (non-quantum) level.

\section{Apparent Luminosity of SNIa}

Let us now discuss the expression for the apparent luminosity of a type Ia supernova in the two cosmological models, SC and CC.

We begin with standard cosmology. Here the number of photons $\frac{{\rm{d}}N}{{\rm{d}}t_{em}}$ emitted per unit physical time by the SNIa and the number of photons $\frac{{\rm{d}}N}{{\rm{d}}t_{obs}}$ received on the earth (per unit time) are related to each other by the redshift factor:
\begin{equation}
\frac{{\rm{d}}N}{{\rm{d}}t_{obs}} = a(z)\, \frac{{\rm{d}}N}{{\rm{d}}t_{em}}=\frac{1}{1+z} \frac{{\rm{d}}N}{{\rm{d}}t_{em}}.
\end{equation}
It is necessary to take into account the redshift effect for the detected photons, so one can obtain that the energy detected per unit time is:
\begin{equation}
\frac{{\rm{d}}E}{{\rm{d}}t} = \frac{1}{1+z}\frac{1}{1+z} L.
\end{equation}
Here $L$ is the absolute luminosity of the SNIa. Having this it is easy to obtain an expression for the apparent luminosity (per unit area):\cite{Weinberg_2008}
\begin{equation}
\ell=\frac{1}{S}\frac{{\rm{d}}E}{{\rm{d}}t} = \frac{1}{1+z}\frac{1}{1+z} \frac{L}{4\pi r^2(z)} = \frac{L}{4\pi d_f^2},
\end{equation}
where the luminosity distance is $d_f=(1+z)r(z)$. In the SC, the Universe expands and $r(z)=(1+z)r$, where $r$ is the comoving distance, which by definition does not depend on the redshift. Therefore, in the Einstein frame:
\begin{equation}
\ell_E =(1+z)^{-4} \frac{L}{4\pi r^2}.
\end{equation}
Let us conduct the corresponding considerations for the conformal cosmological model CC and compare with the results of SC. Note that the number of photons emitted per unit conformal time and absorbed per unit conformal time by the detector is the same:
\begin{equation}
\frac{{\rm{d}}N}{{\rm{d}}\eta_{obs}} = \frac{{\rm{d}}N}{{\rm{d}}\eta_{em}}.
\end{equation}
Incidentally, this statement is valid both in SC and in CC. The physical time elapsed between the emission and absorption of photons differs in the schemes under consideration:
\begin{equation}
\frac{{\rm{d}}N}{{\rm{d}}t_{obs}} =\frac{1}{1+z} \frac{{\rm{d}}N}{{\rm{d}}t_{em}}.
\end{equation}
Thus, both in SC and in CC the energy detected per unit physical time is the same and equal to:
\begin{equation}
\frac{{\rm{d}}E}{{\rm{d}}t} = \frac{1}{1+z}\frac{1}{1+z} L.
\end{equation}
At first glance, in the conformal case $r(z)=r$, and in the Jordan frame a different expression is obtained for the apparent luminosity:
\begin{equation}
\ell_J =(1+z)^{-2} \frac{L}{4\pi r^2}.
\end{equation}
However, it is necessary to take into account that in the conformal frame, the supernova can no longer be treated as an unmodified standard candle. In many respects, this situation is analogous to the redshift effect considered above. The fundamental physics at different points in the Universe has not changed after the transition to the conformal frame (CC); rather, a local change in the units of measurement has taken place. This is evident in the case of cosmological redshift, where precisely this local change in units entails a dependence of the electron mass (and, consequently, of atomic energy levels) on zz
z. In a similar manner, the absolute luminosity of a supernova, LL
L, acquires a conformal weight. Although the precise microphysical conditions of supernova explosions remain an active area of research, their characteristic energy scales (such as the Chandrasekhar limit in standard Type Ia models) are inherently tied to dimensional constants. Thus, while the underlying physics of supernova formation remains invariant, the local units of measurement for all dimensional physical quantities governing the explosion have altered.

Of course, one can thoroughly determine the origin of each redshift factor in certain expressions of the CC model. However, as we have seen earlier, it is much simpler to use the invariant (\ref{invarCC}) and express all observables in units of the reduced Planck mass of the considered frame. Then the corresponding conformal weights of the observables in CC can be calculated on the basis of a simple dimensional analysis. This fact greatly simplifies the analysis of the predictions of the conformal cosmological model and makes the Jordan frame more understandable from an intuitive point of view.

Returning to the question of the apparent luminosity of SNIa within the framework of CC, we obtain that, since its absolute luminosity measured in the corresponding Planck units scales as:
\begin{equation}
[L]=\frac{[E]}{[T]}= L^2 T^{-3} M^1 =M^{2} \,\Rightarrow L_J= a(z)^2 L_E,
\end{equation}
then
\begin{equation}
L_J=\frac{L_E}{(1+z)^{2}}.
\end{equation}
This is precisely the factor that restores the classical equivalence between the SC and CC frames for the case we have explored. Naturally, one can arrive at the same conclusion by considering the characteristic mass scales of the explosion—such as the Chandrasekhar limit in standard Type Ia models—and systematically tracing the emergence of each redshift factor $(1+z)^{\gamma}$ in the CC frame.

\section{Conclusion}

In this work, a comparative analysis of the apparent luminosity of type Ia supernovae was carried out within the framework of the standard and conformal cosmological models. We have shown that the apparent differences in the observational predictions discussed in a number of previous studies arise from differences in the interpretation of physical quantities upon transformation between the Einstein and Jordan frames.

The key result of the study is the establishment of the fact that in the conformal picture type Ia supernovae cease to be "standard candles" in the usual sense of the word: their absolute luminosity acquires a dependence on the redshift as a consequence of the change in local units of measurement. When this effect is correctly taken into account ---which is most transparently analyzed through the dimensional analysis of conformal weights --- the classical equivalence of the results in the SC and CC models is fully restored. Thus, the conformal cosmological model represents not an alternative physical theory, but rather a different, equivalent (at least at the classical level) interpretation of the same cosmological phenomena.

\section*{Acknowledgments}
The authors are grateful to B.N.~Latosh for useful discussions.

\bibliographystyle{elsarticle-num}
\bibliography{SNeIa_eng_arxiv}

\end{document}